V.I.Krystob, L.A.Apresyan, D.V.Vlasov, T.D.Vlasova

# Physical-chemical and chemical modification of PVC and their effect on the electrical properties of the polymer

General Physics Institute RAS
119991 Moscow, Russia

ABSTRACT

The possibilities to increase the static conductivity of PVC compared to the initial antistatic level, for PVC plasticate obtained with effective plasticizer - a so called modifier of type A, and partial thermal dehydrochlorination for formation of a copolymer with a system of conjugated double bonds are investigated. To analyze the dependence of electrical conductivity from double conjugated bonds the doping method is proposed, in which a chemically active dopant is embedded between two thin films of the polymer sample, not touching, and, accordingly, not interacting with the metal electrodes. In this scheme it is found, that the for folding of two layers of partly dehydrochlorinated PVC on the border between them the barrier layer is formed, which can be eliminated by doping with iodine giving significant increase in electrical conductivity. This layer does not arise in the case of type A modifier plasticized PVC. The presence of such layer can be interpreted as a sign of the near-surface spatial heterogeneity associated with the presence of conduction mechanism of double conjugated bonds.

## ВВЕДЕНИЕ

Одним из наиболее известных и распространенных способов модификации полимеров, существенно влияющих на изменение комплекса их физико-механических и электрофизических свойств является введение в их объем ингредиентов различных типов : пластификаторов, наполнителей, стабилизаторов и т.д. Все это в полной мере относится и к поливинилхлориду (ПВХ), одному из наиболее массовых полимеров – диэлектриков, широко используемому практически во всех отраслях промышленности. При этом, если введение различных ингредиентов не приводит к их химическому взаимодействию с макромолекулами ПВХ, т.е. не затрагивает их химического состояния, то можно говорить о модификации ПВХ, носящей физико-химический характер [1,2] .

Типичным случаем подобной физико-химической модификации можно считать использование в качестве ингредиентов-модификаторов различных типов пластификаторов. Например, хорошо известно, что введение пластификаторов в полимер ухудшает диэлектрические характеристики последних, т.е. увеличивает проводимость. Так использование в ПВХ традиционных типов пластификаторов (фталатных, фосфатных и т.д.) способно например, монотонно понижать удельное электрическое сопротивление ($\rho_v$) ПВХ в интервале от $10^{12-16}$ до $10^{9-11}$ Ом·см [2], но по прежнему оставляя его в разряде диэлектриков. Это связано с тем, что введение традиционных типов пластификаторов в полимерную матрицу имеет свои ограничения из-за способности их выделения («выпотевания») в свободном виде на поверхности полимера при достижении порога их насыщения со всеми вытекающими из этого последствиями.

Однако использование новых типов пластификаторов (модификаторов типа А) позволило существенно расширить возможности физико-химической модификации ПВХ не только на количественном, но и качественном уровнях одновременно. Так удельное объемное сопротивление в ПВХ теперь способно было изменяться в более широких пределах (на 8-9 порядков), достигая $\rho_v \sim 10^6$ Ом·см [3] и, таким образом, формально переводить ПВХ из класса диэлектриков в класс антистатиков (полупроводников).

Одновременно при этом удалось зафиксировать совершенно новые и достаточно перспективные электрофизические свойства и аномалии в таких ПВХ-пластикатах [4-6], связанные с различными механизмами переключения электропроводности на четыре порядка величины.

С другой стороны, сегодня известны электропроводящие полимеры, которые уже по самой своей природе обладают достаточно высокими показателями электропроводности, т.е. в которых электропроводностью обладают сами макромолекулы [1,7]. И безусловными лидерами среди них нужно признать полимеры с системой сопряженных двойных связей в цепи (ПСС). Из них наиболее известны сегодня полипиррол, полипарафенилен, политиофен, полиацетилен и т.д., обладающие, в отличие от обычных органических веществ, дополнительно такими качественно новыми свойствами, как наличие парамагнитных центров, полупроводниковыми свойствами, фотоэлектрической чувствительностью и т.д. [1,7, 9]. Для ПСС удлинение участков сопряжения должно приводить к увеличению проводимости как за счет уменьшения ширины запрещенной зоны, так и за счет уменьшения числа межмолекулярных барьеров, преодолеваемых носителями тока под внешним воздействием электрического поля [1, 9]. Следует отметить, что в ПСС в основной цепи концентрация областей сопряжения и их связность выше, чем в полимерах, где основная цепь состоит из насыщенных связей и локализованных сопряженных фрагментов. По этой причине полимеры первой группы могут иметь металлическую проводимость, тогда как во втором случае являются полупроводниками.

Очень часто для усиления электропроводящих свойств полимеров, содержащих ПСС, их подвергают химической модификации с использованием реакций полимера с донорами или акцепторами электронов (допированию), приводящей зачастую к чрезвычайно масштабным изменениям электропроводности полимера. Так, например, допирование полиацетилена электронно-донорными или электронно-акцепторными соединениями приводит к самой высокой по сравнению с другими полимерами электропроводности и фактическому переходу ее в металлическое состояние [1]. При этом следует учесть, что при достижении металлической электропроводности допированного полиацетилена последний обладает малой технологической пригодностью для промышленного использования. Наоборот, при относительно низкой (по сравнению с металлами) электропроводностью модифицированного ПВХ, его комплекс физико-механических и эксплуатационных свойств достаточно просто регулируется физико-химически в широких пределах, что и обеспечивает широкий диапазон применений в различных областях и технологиях.

В свою очередь, исследования проводимости ПВХ с использованием гетерофазных электропроводящих наполнителей (металлических частиц, различных видов сажи и т.д.) [1,2] также имеет свои ограничения, в первую очередь по причине катастрофического ухудшения комплекса технологических и физико-механических свойств ПВХ при увеличении степени его наполнения подобными наполнителями. Поэтому на сегодня достичь оптимума максимально возможных электропроводящих свойств с сохранением при этом необходимого уровня технологических и физико-механических свойств в полимере, на наш взгляд, возможно, либо путем получения полимерных нанокомпозитов с использование различных типов УНТ в качестве гетерофазного наполнителя с концентрацией не превышающей 1-2%, либо с использованием сополимеров, содержащих в своем составе ПСС (например сополимер ПВХ с полиацетиленом) ( ПАц) При этом в идеале получаемый сополимер мог бы обладать проводимостью полимеров, содержащих ПСС, и в то же время комплексом физико-механических свойств модифицированного ПВХ. Тем более, в полимерах с насыщенными связями в основной цепи (типа ПВХ) может иметь место «прыжковая» проводимость при условии , что в структуре полимера имеются фрагменты ПСС [1].

С учетом вышеизложенного было решено изготовить образцы сополимеров полиацетилена и ПВХ и сравнить их с образцами ПВХ, пластифицированных модификатором типа А.

## ЭКСПЕРИМЕНТАЛЬНАЯ ЧАСТЬ.

На первой стадии получения образцов сополимера ПАц-ПВХ осуществляли путем химической модификации ПВХ путем его термичекого дегидрохлорирования из 4%-го раствора в ацетофеноне при T=190$^0$C в течение 11 часов. О прохождении химической модификации (дегидрохлорирования) судили как по изменению окраски самого раствора от бесцветной до красно-коричневой [8] , а также по выделению газообразного хлористого водорода по изменению окраски универсальной индикаторной бумаги. После удаления растворителя (ацетофенона) в термошкафу при T=95$^0$C в течение 24 часов образцы сополимеров ПАц-ПВХ растворяли в тетрагидрофуране (ТГФ), и высушивали в термошкафу при T=45$^0$C 20 часов. Были получены прозрачные слабо окрашенные пленки толщиной примерно 20 мкм. В дальнейшем часть образцов сополимера ПАц-ПВХ подвергалась дополнительной химической модификации (допированию) 5%-ным раствором йода в этиловом спирте. В результате были получены образцы прозрачных пленок с более интенсивной желтой окраской.

Изготовление образцов сравнения осуществляли методом полива на стекло 4 %-го раствора ПВХ в тетрагидрофуране, содержащим модификатор А в массовом соотношении ПВХ:модификатор А= 100:25, соответственно. Указанный модификатор представляет собой смесь диэфиров гликолей (с молекулярной массой 100-400) и монокарбоновых кислот (фракции $C_5$-$C_{10}$), относится к классу эфиров гликолей и монокарбоновых кислот наряду с наиболее распространенными традиционными пластификаторами для ПВХ (ДОФ, который относится к классу эфиров ароматических кислот) и прошел предварительные испытания [3, 9,10].

Высушивание образцов (удаление растворителя) осуществляли в термошкафу при температуре 45$^o$C. В итоге были получены прозрачные ПВХ пленки толщиной примерно 20 мкм. Измерения образцов ПВХ-пленок по показаниям электропроводности осуществляли на приборе, описанном в [4-6 ] с Гостированной круговой измерительной ячейкой. Технологическая пригодность лабораторных образцов оценивалась по удобству работы с ними как при изготовлении, так и при измерениях на приборах.

## ОБСУЖДЕНИЕ РЕЗУЛЬТАТОВ

Данные по измерению электропроводности образцов ПВХ, пластифицированных модификатором типа А, а также образцов ПВХ, подвергшихся операции частичного дегидрохлорирования с обработкой в дальнейшем внутренних поверхностей сложенных двое исходных образцов 5% спиртовым раствором йода, представлены на Рис 1 и Рис.2.

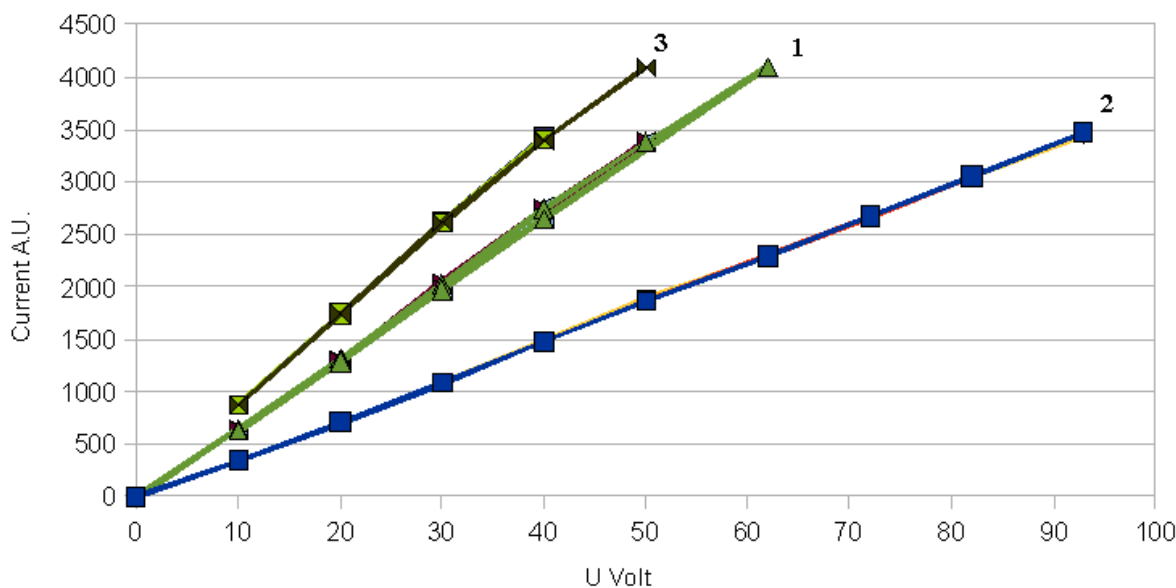

Рис.1. Вольтамперные характеристики образца ПВХ, пластифицированного модификатором типа А в соотношении ПВХ: модюА=100:25 вес.ч., соответственно. (1)- одинарный слой, (2) - двойной (сложенный) слой, (3)- двойной слой с обработкой его внутренних поверхностей 5% спиртовым раствором йода.

Из данных на Рис.1 следует, что при сложении образца вдвое электропроводность сложенного образца (2) по сравнению с одинарным образцом (1) предсказуемо уменьшается примерно в два раза (в рамках закона Ома, пропорционально толщине образца, что свидетельствует о наличии хорошего электрического контакта между слоями). Из данных же Рис.2 для дегидрохлорированного образца электропроводность падает значительно сильнее, практически дол нуля, что позволяет говорить о появлении в образце (2) на границе между поверхностями «запирающего» слоя.

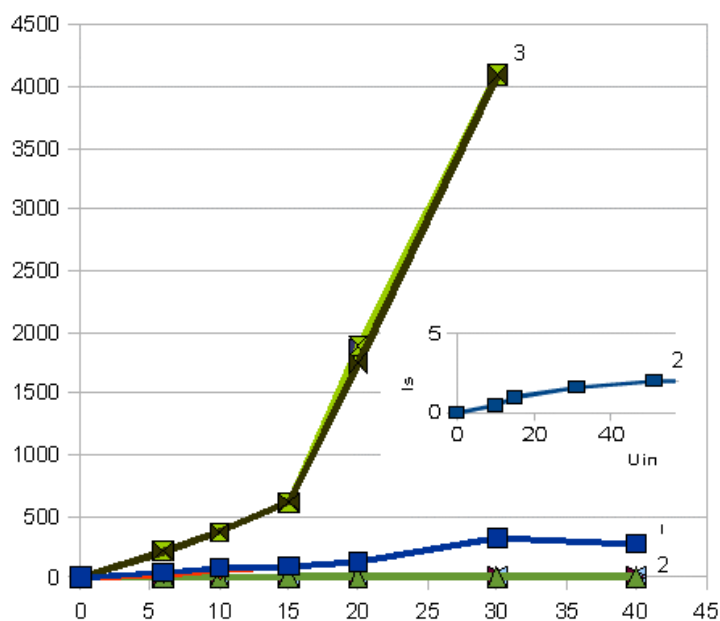

Рис.2. Вольтамперные характеристики дегидрохлорированного ПВХ. (1) - одинарный слой, (2)- двойной(сложенный) слой (на вкладке- в увеличенном масштабе), (3) - двойной слой с обработкой (допированием) его внутренних поверхностей 5% спиртовым раствором йода.

При обработке внутренних поверхностей сложенных вдвое пленок пластифицированного модификатором А образца ПВХ 5% спиртовым раствором йода наблюдается незначительное (примерно в 1,5 раз) увеличение электропроводности (кривые (2) и (3) на Рис. 1). При подобной же обработке внутренних поверхностей дегидрохлорированного образца ПВХ за счет допирования йодом наблюдается значительно более сильное увеличение электропроводности (на 4-5 порядков), а также возникновение сильной нелинейности (см. (2) и(3) на Рис.2). Из этого сравнения следует , что в дегидрохлорированном образце ПВХ происходя значительно более резкие по сравнению с подобными образцами пластифицированного модификатором типа А ПВХ изменения электропроводности как при простом сложении слоя вдвое, так и ее увеличение при допировании йодом.

Таким образом, с учетом того, что частичное дегидрохлорированные образцы ПВХ обладали дополнительно при этом высокой технологичностью и необходимым комплексом эксплуатационных свойств (незалипаемость, удобное отделение от подложки, возможность многократного измерения без ухудшения внешнего вида, органолептических свойств, отсутствием следов коррозии на измерительных электродах и т.д.), можно сделать вывод о том, что совместное использование как дегидрохлорирования, так и модификатора А, открывают принципиальную возможность регулирования и оптимизирования электрофизических и физико-механических свойств ПВХ на количественном и качественном уровне одновременно.

СПИСОК ЛИТЕРАТУРЫ

1. Блайт, Д. Блур. *Электрические свойства полимеров*: М.: Физматлит-2008г.

2. Энциклопедия Полимеров. Ред. коллегия: В. А. Кабанов (глав. ред. )[и др. ] Т.2. М., Сов. Энц. , 1974. 1224 с.

3. V.I. Kryshtob. *Method for production of antistatic polymer materials*. USA Patent N 5.576.383 from 19.11.1996 .

4. D. V. Vlasov, L. A.Apresyan, T. V. Vlasova, and V. I.Kryshtob - *Nonlinear Response and Two Stable Electroconducting States* in Transparent Plasticized PVC Films.Technical Physics Letters, 2009, Vol. 35, No. 10, pp. 923–925.

5. D.V. Vlasov, L.A. Apresyan, V.I. Krystob, T.V. Vlasova, 2011, *Anomalies and error limits in electrical-conductivity measurements in plasticized transparent poly(vinyl chloride) films*, Polymer Sci., ser.A, 53: 5, 430-436. 2011.

6. D. V. Vlasov, L. A. Apresyan, T. V. Vlasova, V. I. Kryshtob, 2011; *On Anomalies of Electrical Conductivity in Antistatic Plasticized Poly(vinyl chloride) Films,* American Journal of Materials Science, vol. 1(2): 128-132. DOI:10.5923/j.materials.20110102.21

7. А.Г.Гроздов и др. *Электропроводящие полимеры*// Химическая промышленность сегодня, №5, 2007, с.1-5.

8. К.С.Минскер. *Деструкция и стабилизация поливинилхлорида.* Химия, 1972.

9. Д.В.Власов, В.И. Крыштоб, Т.В.Власова, С.Н.Бокова, О.П. Шкарова , Е.Д.Образцова, Л.А. Апресян, В.И.Конов, *Получение композитов с чередующимися слоями поливинилхлорида и одностенных углеродных нанотрубок, однородно диспрегированных в карбоксиметилцеллюлозе* , Высокомолекулярные соединения, Серия А, т. 54, No. 1, (2012)  с. 39–43.

10. S.N. Bokova, D.V. Vlasov, V.I. Kryshtob, T.V. Vlasova, E.D. Obraztsova, L. A. Apresyan, V.I. Konov, *Laminated Composites of PVC and Single Wall Carbon Nanotubes, Dispersed Uniformly in CMC*, Journal of Nanoelectronics and Optoelectronics , Vol. 7, 87–89, 2012.